# Layer-by-Layer Assembled Nanowire Networks Enable Graph Theoretical Design of Multifunctional Coatings

*Wenbing Wu, Alain Kadar, Sang Hyun Lee, Bum Chul Park, Jeffery E. Raymond, Thomas K. Tsotsis, Carlos E. S. Cesnik, Sharon C. Glotzer,\* Valerie Goss,\* Nicholas A. Kotov\**

**Wenbing Wu**
Department of Chemical Engineering; Biointerfaces Institute
University of Michigan
2800 Plymouth Rd, Ann Arbor, MI 48109, USA
**Alain Kadar**
Department of Chemical Engineering; Biointerfaces Institute
University of Michigan
2800 Plymouth Rd, Ann Arbor, MI 48109, USA
**Sang Hyun Lee**
Department of Electrical Engineering and Computer Science; Biointerfaces Institute
University of Michigan
2800 Plymouth Rd, Ann Arbor, MI 48109, USA
**Bum Chul Park**
Department of Chemical Engineering; Biointerfaces Institute
University of Michigan
2800 Plymouth Rd, Ann Arbor, MI 48109, USA
**Jeffery E. Raymond**
Department of Chemical Engineering; Biointerfaces Institute
University of Michigan
2800 Plymouth Rd, Ann Arbor, MI 48109, USA
**Thomas K. Tsotsis**
The Boeing Company
14441 Astronautics Dr, Huntington Beach, CA 92647, USA
**Carlos E. S. Cesnik**
Department of Aerospace Engineering
University of Michigan
3000 François-Xavier Bagnoud Aerospace Building
1320 Beal Avenue Ann Arbor, MI 48109-2140
**Sharon C. Glotzer**
Department of Chemical Engineering; Biointerfaces Institute
University of Michigan
2800 Plymouth Rd, Ann Arbor, MI 48109, USA
**Valerie Goss**
Department of Chemistry, Physics & Engineering Studies
Chicago State University
9501 S. King Dr. WSC 234
Chicago, IL 60628
**Nicholas A. Kotov**
Department of Chemical Engineering, Biointerfaces Institute
University of Michigan
2800 Plymouth Rd, Ann Arbor, MI 48109, USA;
Department of Aeronautics, Faculty of Engineering
Imperial College London
South Kensington Campus
London, SW7 2AZ

Corresponding authors: sglotzer@umich.edu; vgoss@csu.edu kotov@umich.edu.

**ABSTRACT:** Multifunctional coatings are central for information, biomedical, transportation and energy technologies. These coatings must possess hard-to-attain properties and be scalable, adaptable, and sustainable, which makes layer-by-layer assembly (LBL) of nanomaterials uniquely suitable for these technologies. What remains largely unexplored is that LBL enables computational methodologies for structural design of these composites. Utilizing silver nanowires (NWs), we develop and validate a graph theoretical (GT) description of their LBL composites. GT successfully describes the multilayer structure with nonrandom disorder and enables simultaneous rapid assessment of several properties of electrical conductivity, electromagnetic transparency, and anisotropy. GT models for property assessment can be rapidly validated due to (1) quasi-2D confinement of NWs and (2) accurate microscopy data for stochastic organization of the NW networks. We finally show that spray-assisted LBL offers direct translation of the GT-based design of composite coatings to additive, scalable manufacturing of drone wings with straightforward extensions to other technologies.

**Keywords:** structural complexity; silver nanowires; layer-by-layer assembly; composite electrode; anisotropic conductivity.

## 1. INTRODUCTION

Advanced coatings and composites from nanomaterials enable current and future space, robotics, transportation, information, biomedical, and energy technologies. While many processes have been used to create nanostructured coatings[1–5], the minimization of process energy, input materials, and waste toxicity is essential for sustainability of emerging technological processes. Cumulatively, these requirements favor processes based on self-assembly at ambient temperature from aqueous dispersions. This has led to renewed interest in the layer-by-layer assembly (LBL)[6–8] due to its eco-friendliness, scalability, additivity, and universality. The ability of LBL to create on-demand, high-performance composites from inexpensive and imperfect components is remarkable[9–12], though in practice it has been largely an empirical endeavor.[13–15]

In addition to these attributes, there exists another advantage to LBL that remains largely untapped. Breakthrough materials must often combine electrical, mechanical, thermal, and optical functionalities. And these properties must be optimized simultaneously. The realistic prediction and optimization of several properties at once require application of several computational models calculating, for instance, mechanical deformations, charge transport, and optical absorption. Such material design protocols are computationally expensive and necessitate teams of researchers with different expertise.[16] While being implemented for atoms and being theoretically possible for nanocomponents, the actual realization of in-silico designed macroscale nanostructured materials with simultaneous optimization of multiple properties remains elusive. This is true even considering the development of complementary protocols using artificial intelligence (AI) aided by constantly expanding computational power. The problem is that the assumptions used in material design and the realities of nanomaterial synthesis present a significant mismatch in terms of nanocomponent polydispersity (*e.g.* size, shape, etc.), variance in their long-range organization and various interfaces incorporating hard-to-describe disorder. For example, typical computational models of nanoscale self-assembly assume component monodispersity, which deviates drastically from the ground reality of sourced, readily available nanomaterials with high polydispersity. Furthermore, systems with highly asymmetric components and dynamic geometries (e.g. nanosheets, NWs, or nanofibers) are simultaneously of high practical significance and computationally problematic.[17]

Four special features of LBL enable effective integration of experimental materials synthesis with a materials-on-demand computational framework. *First*, LBL confines nanoscale components to quasi two-dimensional (2D) adsorption planes. The associated reduction of the degrees of geometrical freedom for nanocomponents in Cartesian space makes computational representations much faster and thus more accurate. *Second*, nanoscale 2D imaging is far more accurate and accessible than 3D techniques (e.g., tomography). This feature allows for rapid validation of the computational models, which is essential for property predictions. *Third*, being suitable to curvilinear surfaces, LBL offers a simple and scalable path to 3D macroscale materials. *Fourth*, LBL is applicable to an exceptionally wide range of nanomaterials, polymers, and biological structures with different shapes, sizes, and compositions.

These four unique characteristics are expected to permit new, computationally efficient methodologies for modeling, design, property prediction and structural optimization. In this work, we show that indeed LBL affords the accurate design and practical macroscale implementation of composite coatings through the use of Graph Theory (GT), the emerging approach to materials design capable of bridging different scales and easily integrable with AI techniques.[18] The GT approach uses graphs extracted from experimental microscopy to computationally construct any nanoscale architecture that may be synthesized with LBL. We specifically show that spray-assisted LBL[19–24] can be used to produce stochastically organized composites combining both order and disorder from silver NWs. We elucidated their electrical and terahertz (THz) optical properties. GT models were able to predict the conductivity, failure currents, and optical anisotropy of the experimental NW composites. The direct implementation of this process for macroscale composite coatings is demonstrated for robotics and transportation technologies (i.e., drones).

## 2. RESULTS AND DISCUSSION

### 2.1. Graph Theoretical Models and Related Validation Problems

Graphs are sets of nodes, $n$, and edges, $e$. All self-assembled nanostructures can be described as graphs following a simple convention based on minimal representations of nanostructures.[18] Composites utilizing NWs, nanotubes, nanofibers, nanoribbons, etc., can be generally referred to as $K_2$-nanocomponents based on the node-edge-node $K_2$ complete graphs that represent rod-like nanocomponents.[18] For continuous networks of $K_2$-nanomaterials, intersections are represented as nodes and NW segments connecting them are represented by edges. The conceptual advantage of GT over other models is that GT (1) naturally incorporates both order and disorder in these materials; (2) facilitates assessment of properties related to the connectivity patterns as exemplified by charge, heat, and stress transfer; and (3) enables the application of network theory to scale-up small experimental assemblies to arbitrarily large nanoscale systems.[25–27]

Computer simulations have demonstrated that mechanical[28], electrical[29], and biological[30] properties of nanoscale materials could be related to GT parameters (also known as indices or descriptors) of percolating networks. The theoretical possibility of GT methodologies for the combined assessment of electric and mechanical properties of carbon nanotube networks were pointed out, though they remain unexplored.[31] Prior examples of GT applied to complex particles[18] and branched polymeric composites[32] also show that property assessment based on GT parameters reduces computational cost by 10-1000 times.

However, the relationships between GT structural parameters and physical properties of nanoscale materials remain hypothetical without experimental validation.[33–35] In this regard, the validation of electrical conductivity models is of central concern. This methodological gap is stunning because there is no lack of conductive $K_2$ nanocomponents. The problem is that an adequate representation of a macroscale network from nanoscale rod-like elements requires angstrom-scale imaging accuracy of their junctions. This is due to a need to define the absence of nodes, $n$, because electrical conductivity through a 3D network (as well as electromagnetic resonances and heat/stress transfer) are dependent on these small gaps.[36] More generally, a change in the distance between the NWs, often 10,000 - 1,000,000 times smaller than the characteristic dimensions of the test sample, makes a big difference. In the case of standard composites and coatings, NWs are stochastically distributed in a 3D space and can be bent in multiple directions. Because of this, imaging of the entire 3D volume with such accuracy is hardly possible, even with the latest advances in electron microscopy tomography.[37]

Composites and coatings made by LBL of NWs are different because their accurate GT representation *can* be constructed using standard 2D imaging techniques, such as scanning electron microscopy (SEM) and atomic force microscopy (AFM). When composite construction occurs from sequential layering, the degrees of freedom for NWs from each bilayer are restricted. Taking NWs from silver as an example (**Figure 2a – d**), an *adsorb-rinse-adsorb-rinse...* assembly approach forces NWs to lie in the close proximity while the rinsing step removes extra polymer that may increase the gaps between NWs. Thus, every NW-NW junction observed by SEM or AFM images of single layer films can be treated as an electrically conductive junction resulting in the confident placement of a node. Furthermore, the open source software package, *StructuralGT*[32], offers automated extraction of GT parameters from SEM and AFM images. This then enables the direct input of GT results into the computational construction of multilayer coatings and property predictions.

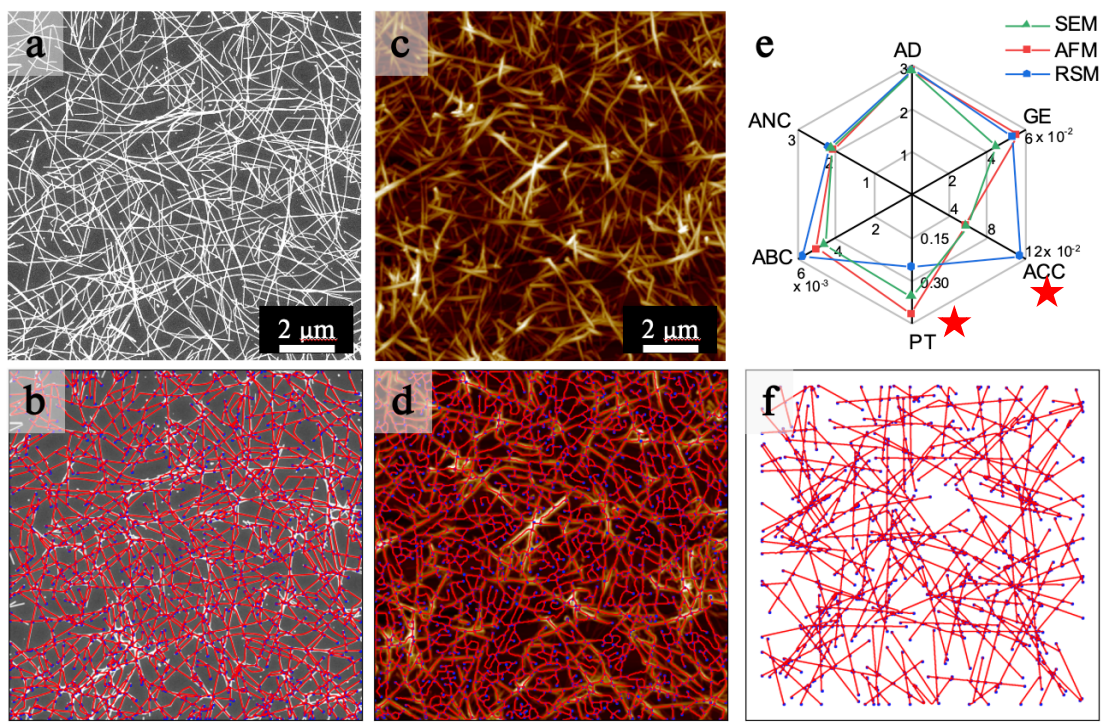

**Figure 1.** SEM image (**a**) and AFM image (**c**) of a single-bilayer film and their GT representations (**b** and **d**, respectively) obtained by *StructuralGT*. (**e**) Selected GT parameters obtained from the SEM/AFM images and a computational random stick model (RSM): average degree, AD; global efficiency, GE; average clustering coefficient, ACC; percolation threshold, PT; average betweenness centrality, ABC; average nodal connectivity, ANC. GT parameters are average values over four samples. (**f**) GT representation of a simulated RSM model for a NW network.

GT analysis of AFM and SEM images demonstrates that the conventional simulation models used for $K_2$ nanostructures only partially capture the organization of the coatings, even assuming that all the nodes and junctions are assigned correctly. One can compare, for instance, GT parameters generated by a conventional random stick model[35] (RSM, **Figure 1f, 1e**, SI) and those extracted from SEM (**Figure 1a** and **1b**) and AFM (**Figure 1c** and **1d**) images. The RSM simulation yields an average clustering coefficient, ACC, that is twice as high as the experimental results. RSM simulation also presented an increased average betweenness centrality (ABC) and a reduced percolation threshold (PT). The ACC discrepancy indicates a higher number of intersection clusters in RSMs, arising because RSM does not account for the excluded volume effects inherent in real-world materials. Increased ACC and reduced PT are both strong indicators that RSMs underestimate sheet resilience, likely due to the structural heterogeneity highlighted by increased ACC.

**2.2. Isotropic and Anisotropic Nanowire Networks by made by LBL Deposition**

Alternating spray deposition of polyethyleneimine (PEI) and silver NWs (Figure S1a), we obtained layered composites with a cumulative structure of (PEI/NWs)$_N$ that lack a theoretical upper limit for $N$.

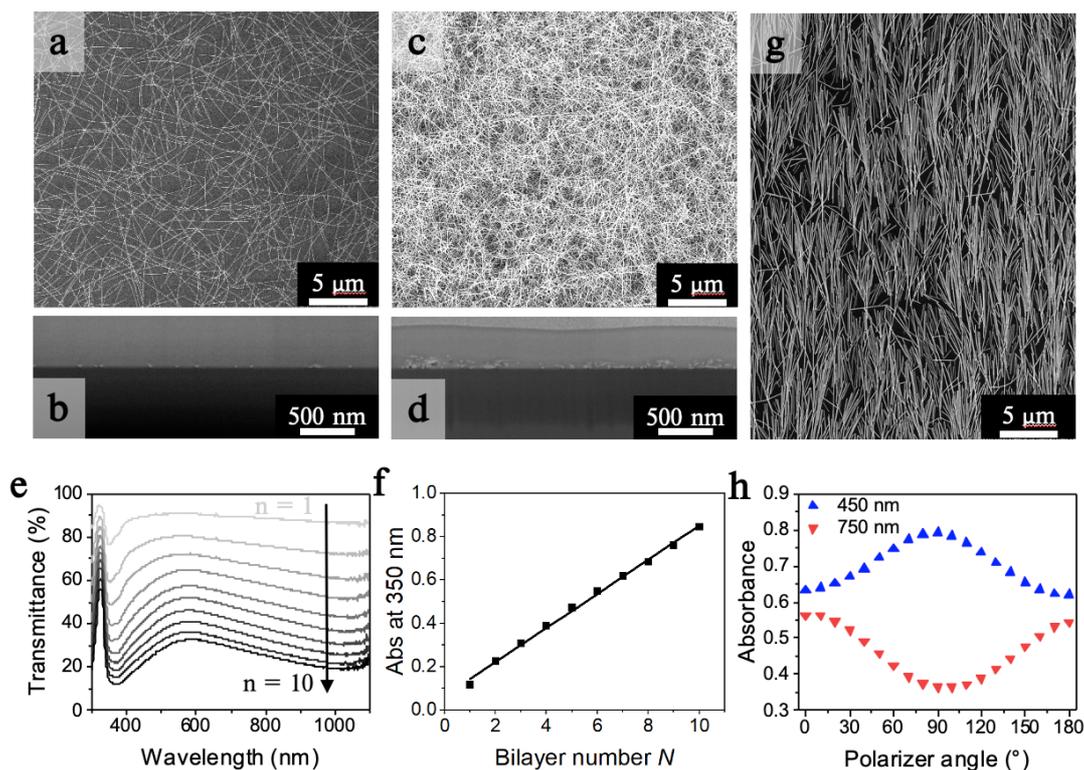

**Figure 2.** Top-view (**a**) and cross-sectional (**b**) SEM images of the single-bilayer film. Top-view (**c**) and cross-sectional (**d**) SEM images of the 10-bilayer film. (**e**) Transmittance spectra of the (PEI/NWs)$_N$ films ($N = 1 - 10$). (**f**) Absorbance at 350 nm of the (PEI/NWs)$_N$ films as a function of bilayer number N. (**g**) SEM image of the aligned single-bilayer film. (**h**) Absorbance at 450 nm and 750 nm of the aligned single-bilayer film as a function of the polarizer angle.

SEM images of single-bilayer ($N = 1$) and 10-bilayer ($N = 10$) NW films, top-down and cross-sectional, are presented in **Figure 2a–d** and Figure S2. The single-bilayer coating covers 15% of the surface area of the glass substrate. The surface coverage increased to 84% for $N = 10$ and the thickness was measured to be 93 nm (**Figure 2d**). (PEI/NWs)$_N$ multilayers with $N = 1 - 10$ have high transmittance in the UV-Vis-NIR (**Figure 2**), with T% = 91% at 550 nm. The linear increase in absorbance of the films with $N$ confirms the reproducibility of each layer and consistently retained 2D NW confinement (**Figure 2f**).

The root-mean-square (RMS) roughness of the single-bilayer sample was 32 nm (Figure S3), similar to the NW diameter. The roughness increased with $N$ (Figure S3f) non-linearly and approached an asymptotic value of 63 nm. The network is quantitatively demonstrated to be isotropic and stochastic, but with consistent *N*-independent organizational pattern (Table S3).

Anisotropic films (**Figure 2g**) were obtained by grazing incidence spraying, resulting in NW alignment (Figure S1b). Excitation of the transverse and longitudinal plasmonic modes of NWs depends on the angle α between the long axes of NWs orientation and the axis of light polarization incident on the coating. When α = 0° (i.e., photons electric field oscillations aligned with NW orientation), the longitudinal modes are preferentially activated; at α = 180° the transverse modes are preferentially activated. The absorption of a polarized source (Figure S4a) at a short wavelength (450 nm) increased from 0° to 90° and decreased from 90° to 180°; the opposite was observed at a longer wavelength (750 nm) (**Figure 2h**). The isotropic films had unchanged absorption spectra at different polarization angles (Figure S4b), serving as both a control and a strong indicator of the alignment of the grazing angle-produced films.

**2.3. Multifunctional properties of the LBL Films**

The NW LBL films combine high conductivity, low density, high thermal resilience, and THz electromagnetic resonances. In this work, we primarily focus on electrical and electromagnetic properties because the problems caused by the mismatch in scales, specifically for tunneling junctions and the macroscale implementations of conductive coatings, is the highest.

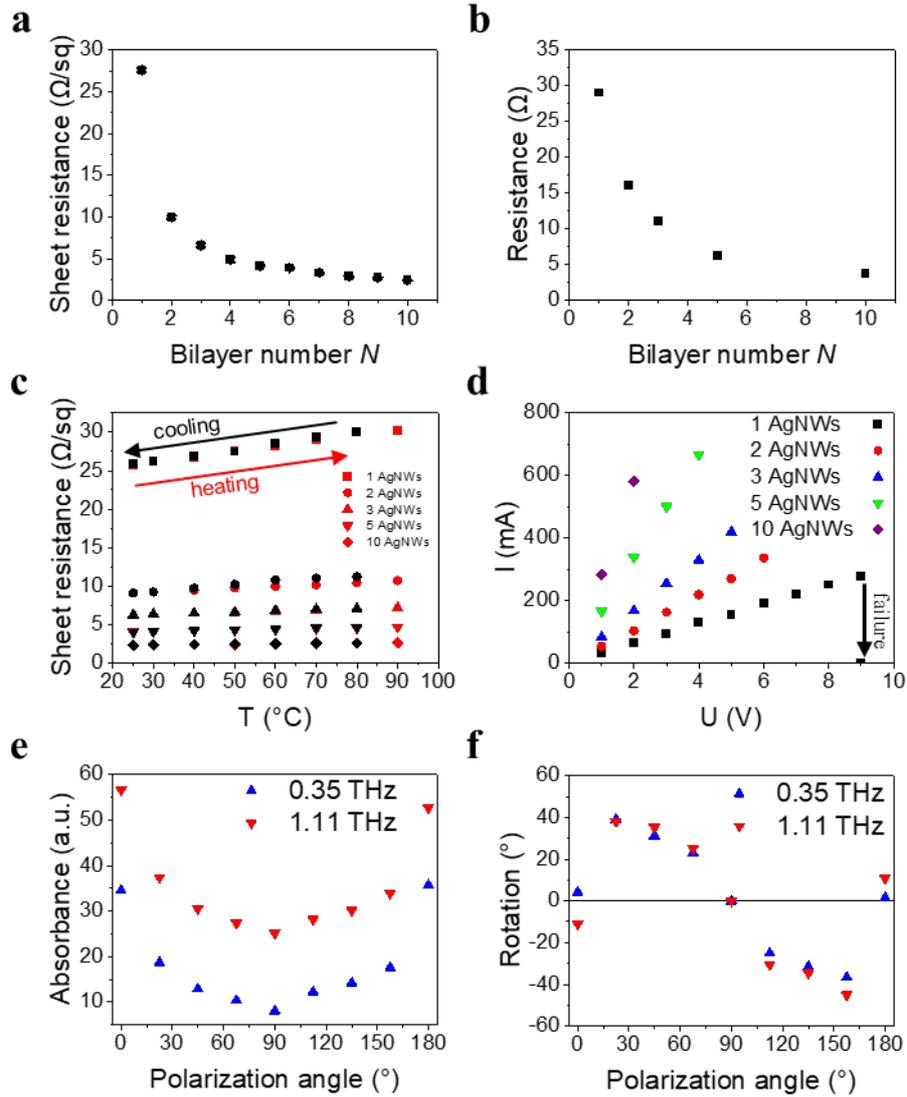

**Figure 3.** (**a**) Sheet resistance of the (PEI/NWs)$_N$ films deposited on glass slides as a function of bilayer number *N*. (**b**) Resistance of the (PEI/NWs)$_N$ films deposited on Kapton substrates as a function of *N.* (**c**) Variation of sheet resistance as a function of temperature through heating and cooling of (PEI/NWs)$_N$ films deposited on glass slides. (**d**) *I-U* curves of (PEI/NWs)$_N$ films deposited on Kapton substrates until sample failure. Absorption spectra (**e**) and optical rotation (**f**) at 0.35 THz and 1.1 THz as a function of polarization angle of aligned NW films.

The single-bilayer and 10-bilayer films possessed sheet resistances of 28 and 2 Ω sq$^{-1}$, respectively, with an approximately reciprocal trend between *N* and sheet resistance. (**Figure 3a**) This sheet resistance is remarkably lower than what can be achieved with ITO glasses (typically 10 – 100 Ω sq$^{-1}$). Such a low sheet resistance is attributed to surface confinement of NWs during the deposition, which promotes an exceptionally high interconnectivity of NWs. Although we use *N* = 10 in this study, there is no limit on the number of bilayers that can be produced. The area density of the $N = 1$ film is calculated to be $4.72 \times 10^{-5}$ kg m$^{-2}$, which demonstrates its appropriateness for use as a lightweight and highly conductive film in, for example, aerospace applications. The films were also found to have excellent thermostability, tested over a temperature range of 25°C to 90°C. The sheet resistance of the films increased with temperature, with no permanent changes to conductivity with thermal cycling (*i.e.,* full recovery, **Figure 3c**). Sheet resistance increased by 18% on average at 90 °C relative to 25 °C, which is reasonable given that a 14% increase is predicted for this range in individual silver NWs.[38] Samples were also tested after exposure to low-temperature (-196 °C in liquid nitrogen) and vacuum conditions. Subsequent measurements at ambient temperatures and pressures demonstrate that changes in conductivity were negligible.

Spray-assisted LBL also allows for a wide array of substrate choices, including flexible or curvilinear surfaces. (PEI/NWs)$_N$ multilayers were deposited on Kapton sheets, common to electrical circuits used in aviation and wearable/flexible devices. Applying a DC voltage at the ends of 2 cm × 2 cm samples, *I-U* curves were measured (Figure S5a) for the (PEI/NWs)$_N$ films (**Figure 3b**). Like films on glass substrate, a reciprocal trend between bilayer number and the resistance was observed. The values of the resistance (Ω) measured on square-shaped Kapton substrates are very close to those of the sheet resistance (Ω sq$^{-1}$) measured on glass slides, demonstrating a high reproducibility for LBL films on different substrates.

Current-carrying capacity was measured by gradually increasing DC voltage from $U = 1$ V to charge transport failure (**Figure 3d**), with current being recorded at the onset of bias to avoid Joule heating. The single-bilayer structure failed at $U = 9$ V and $I = 278$ mA. Unsurprisingly, coatings with $N > 1$ carried more current, with failure occuring at lower voltage because of reduced resistance. For $N$ = 5 and 10, exceptionally high current-carrying capacities (~600 mA) were observed.

For aligned films on glass slides, the sheet resistance was 13 Ω sq$^{-1}$ and 24 Ω sq$^{-1}$ in the parallel and transverse directions, respectively. Similar resistances and anisotropies were also found for aligned film deposited on Kapton (Figure S5b).

The NW films exhibited strong optical activity in the THz range (Figure S6). The aligned film showed differential THz absorption (TA) and THz optical rotation dispersion (TORD) dependent on the polarization angle, *i.e.*, the relative angle between incident THz polarization direction and NW alignment direction. The TA decreases from 0° to 90° and increases from 90° to 180° (**Figure 3e**), revealing enhanced THz interactions induced by long-range order of the NWs along the alignment direction. The THz polarization is also found to be rotated towards the NW alignment direction, with a maximum rotation angle up to 45° at 1.1 THz (**Figure 3f**).

Young's modulus of the films was estimated using a measurement of film buckling on soft substrates (Figure S7) [39]. Composite coatings based on ten-bilayer LBL films display a buckling wavelength of 2.82 *μ*m (Figure S7e), calculated to represent a modulus of 879 MPa.

**2.4. Graph Theoretical Models**

GT allows one to establish direct relations between properties and structure with disorder. The organization of nanostructures is stochastic but non-random and follows specific connectivity patterns. GT structural descriptors include clustering coefficient, percolation threshold, eigenvector centrality, graph Laplacian, and others. Bilayer number $N$ determines the total number of nodes and edges, which is also important for macroscale properties. In respect to the properties, i.e. measurable characteristics, we are interested in sheet resistance ($\Omega_{eff}$), anisotropy ($Y_{pred}$), photon absorbance, and current carrying capacity. The results for experimental and predicted sheet resistance in $N = 1$ isotropic films are shown in **Table 1**.

| Batch | Sample | $\Omega_{eff,exp}$* | $\Omega_{eff,pred}$ |
|---|---|---|---|
| $R_J = 7\ \Omega$ | Sparse - 200 s | 72 | 87 ± 8 |
| | Sparse - 150 s | 95 | 98 ± 16 |
| | Sparse - 60 s | 159 | 130 ± 16 |
| | Sparse - 30 s | 244 | 203 ± 62 |
| $R_J = 2.75\ \Omega$ | Dense | 27.6 ± 0.2 | 28 ± 3 |
| n > 5 for $R_J = 7\ \Omega$, n = 30 for $R_J = 2.75\ \Omega$ | | | |
| * relative standard deviations (RSD) for all experimental values <10%. All results in units of $\Omega$ sq$^{-1}$ | | | |

**Table 1:** Experimental and predicted sheet resistances for isotropic conductive films. '$pred$' is calculated effective resistance, '$exp$' is the experimental value. All resistances are in $\Omega$ sq$^{-1}$ and the error intervals are standard deviations about the mean. The sparse samples include spaying time, in seconds.

| Resistance Direction | Glass | | Kapton |
|---|---|---|---|
| | $\Omega_{eff,exp}$ | $\Omega_{eff,pred}$ | $\Omega_{eff,exp}$ |
| Aligned | 25 ± 14 | 15 ± 12 | 12 |
| Transverse | 35 ± 4 | 41 ± 8 | 25 |

**Table 2.** Experimental and predicted aligned and transverse resistances for aligned conductive films. '$pred$' is calculated effective resistance, '$exp$' is the experimental value. Glass resistances are in $\Omega$ sq$^{-1}$ and Kapton resistances are in ohms. The error bars are standard deviations about the mean. Thermostability and current-carrying capacity results for the aligned film can be found in **Figure 3**.

The GT models show that multilayers can be treated as a direct extension of the single bilayer model, as evidenced by the linear increase in absorbance of the films as a function of $N$ (**Figure 2h**). For multilayer models, connections between adjacent points of neighboring bilayers are added according to the procedure described in the SI. **Figure 4b** displays that this methodology results in a better prediction (lower residuals, Table S2) of experimental data than a simple parallel resistor model (SI).

The GT model may also be applied to the films of aligned wires and used to predict anisotropy of the sheet resistance and NW alignment ($Q_{xx}/Q_{yy}$). **Table 2** summarizes the experimental and GT results for glass slides, and experimental results for Kapton substrates. They show that conductivity increases by as much as 100%.

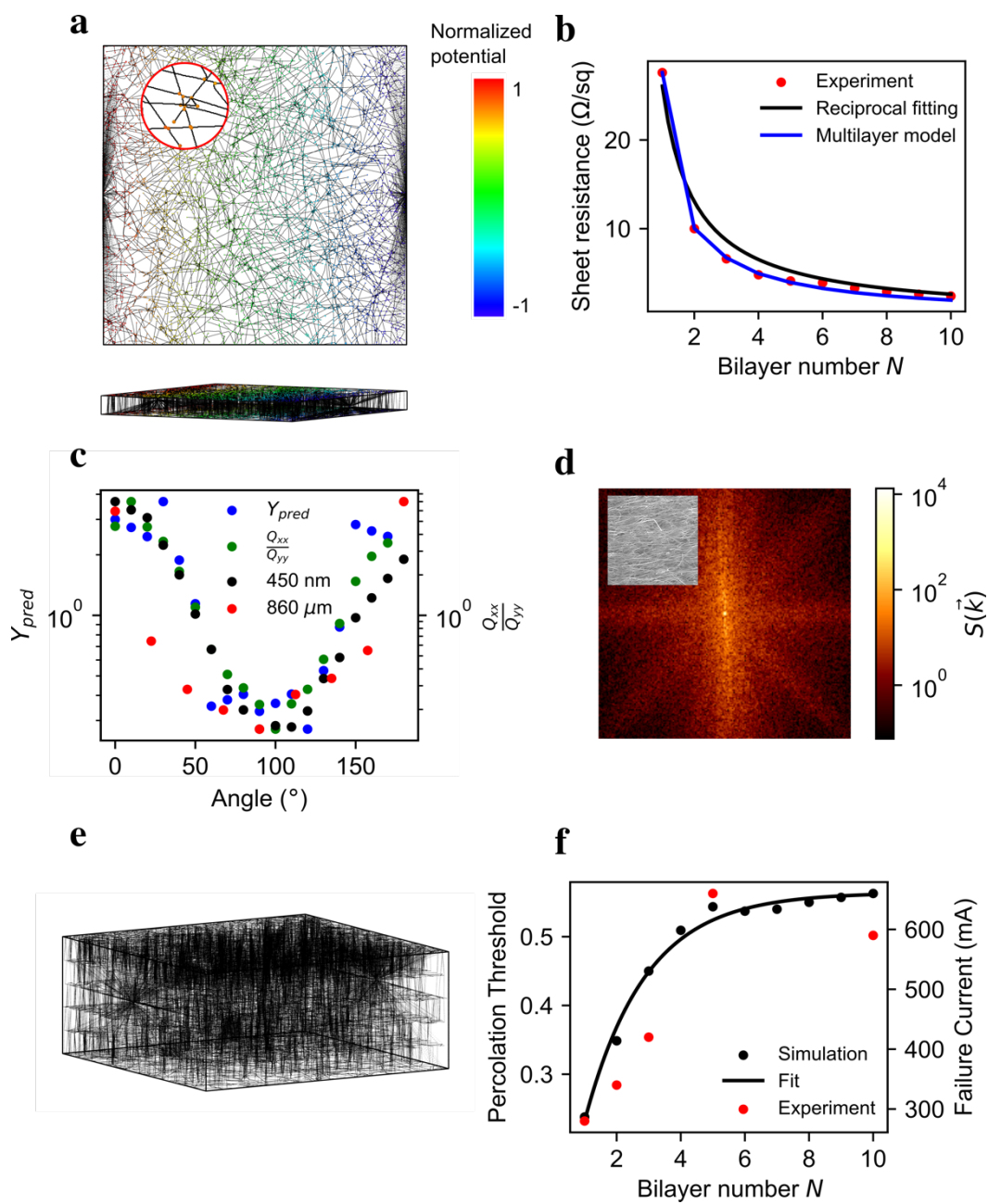

**Figure 4.** (**a**) Schematics of charge potential in the LBL film, $N = 2$ colored by normalized electric potential. The layers have interlayer connections, allowing resistance to drop to 10 Ω sq$^{-1}$. The inset presents upper and lower node connectivity, which facilitates interlayer flow. (**b**) Sheet resistance results for the multilayer model. (**c**) Predicted selectivity, alignment factor, and experimental absorbance as a function of film rotation. Absorbances have been normalized

between 0 and 1. (**d**) Diffraction patterns for the aligned NW film, with SEM inset. (**e**) A coating with $N = 6$ film predicted based on SEMs of $N = 1$ film with interlayer connections. (**f**) Correlation between percolation threshold and failure current.

Importantly, experimentally validated GT models (with nodes associated to specific locations in the Cartesian space) also allow one to investigate the conductivity as a *continuous* function of rotation. In such an analysis, the anisotropy at a particular orientation is best quantified by calculating the directional selectivity, $Y_{pred}$, defined as

$$Y_{pred} \equiv \Omega_{eff}/\Omega_{eff}^{\perp} \qquad \text{Equation. 1}$$

Where $\Omega_{eff}^{\perp}$ is the film's effective resistance when a potential is applied along a direction perpendicular to the direction applied in obtaining $\Omega_{eff}$. **Figure 4c** displays selectivity as a continuous function of rotation for the aligned film. We can demonstrate that the conductive anisotropy comes from the alignment of the wires, by calculating the nematic tensor, $Q$, defined as

$$Q \equiv \langle u_i u_j - 1/3 \, \delta \rangle \qquad \text{Equation. 2}$$

Where $u_i$ is the vector tangent to wire and $\delta$ is the unit tensor. From our definition of $Y_{pred}$, we expect that it should correlate with the ratio of the diagonal elements of $Q$, which we call the alignment factor, $Q_{xx}/Q_{yy}$. Furthermore, given that we know modes of the film's plasmonic resonances correspond to vibrations along the principal axes of the NWs, we expect that both $Y_{pred}$ and $Q_{xx}/Q_{yy}$ should correlate with the film's absorbance. All of these correlations are confirmed and presented in **Figure 4c**.

The scalar nematic order parameter, *P2*, may also be calculated to assess the orientational ordering of the films and validate the findings. For wires confined to 2D layers in 3D space, the value of the scalar nematic parameter at maximum randomness is 0.25 (as shown in the SI). The aligned film has a high *P2* (0.43), indicating alignment of the NWs. The film also displays an anisotropic diffraction pattern (**Figure 4d**).

However, some films may have anisotropic conductivities whilst exhibiting neither anisotropic diffraction patterns nor obvious NW alignment. In these cases, *P2* is especially powerful because it can be used to detect this anisotropy when diffraction and SEM cannot (SI).

Current-carrying capacity of the LBL composites was experimentally measured to obtain a maximum current to failure (**Figure 3d**). Computationally, network failure is an well-studied phenomenon, and can be used to predict the conductivity failures. In GT, this is usually investigated by calculating the percolation threshold, $p$[40]. To construct multilayer networks for *p* calculations, we use the same density of interlayer connections used in the two-bilayer conductivity model to model composites up to $N = 10$. **Figure 4e** shows an example of how SEM images of $N = 1$ films were used to construct a composite coating with $N = 6$ bilayers. To estimate *p*, we repeatedly remove nodes until source and sink become disconnected. **Figure 4f** shows a clear correlation between *p* and failure current. The fit is of the form $y = a(1 - e^{-bx})$, chosen because we expect *p* and failure current to both asymptotically approach a finite value. Hence, *p* is an appropriate non-destructive assessment of a film's current capacity. This finding is uniquely useful in that it allows for the possibility of reverse-engineering films to meet application-specific electronic requirements without extensive trial-and-error.

We note that $\beta$ has previously been used as a metric for network failure[41,42]. However, those works focus on identifying locations of network failure, as opposed to predicting global thresholds. Furthermore, in the SI, we show for our system mean betweenness metric is inversely proportional to *N*. This finding indicates that $\beta$ can be suitable for other electrical characteristics but not for the current carrying capacity.

## 2.5. Conformal Coatings

As highlighted previously, composite coatings from LBL allows deposition of uniform conformal layers on curvilinear surfaces.[43–45] This is especially important for textured, twisted, and/or perforated surfaces typical of advanced materials and 3D printed components. Curvilinear surfaces are of particular interest because they are associated with unique symmetries and anti-symmetries (ex. helices) and low symmetry periodic structures typical of ordered and disordered metamaterials. Furthermore, curvilinear surfaces afford direct conversion of nano- and microscale 2D coatings into macroscale 3D objects. This feature is of critical importance in the context of additive manufacturing, device translation, and scalability.

To demonstrate the possibility of rapid model-to-device translation, deposition of NW composite coatings were carried on 12" × 12" PEI/Kapton substrates (**Figure 5a,b**). The sheet resistance of the sample was measured to be 417 $\Omega$ by two-point measurements (Figure S11). This value is 15 times higher than the 1" × 1" sample with the same structure, appropriate given the 12-fold increase in length. This demonstrated that fast deposition over even larger areas would be possible by implementing spray automation.

We applied spray deposition of NW films to a drone wing made of epoxy and reinforced with carbon fibers (AS4/3506-1), a commonly used composite in the aerospace industry, to demonstrate capability in this area (**Figure 5**). The conductivity of the intrinsically conductive carbon fiber-based part **(Figure 5f)** saw a four-fold increase (**Figure 5i**) after NW coating. This result is remarkable given the nanoscale thickness and negligible weight of the coating, as well as the size of the part.

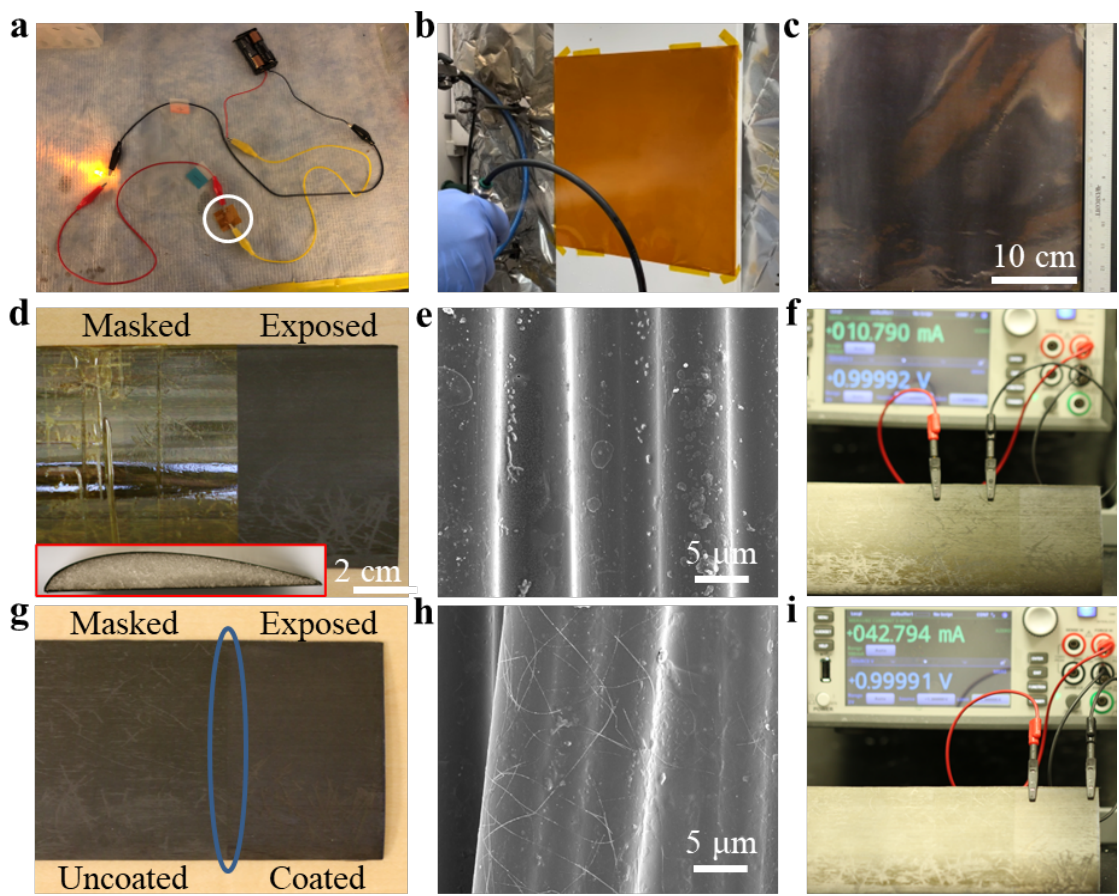

**Figure 5.** (**a**) LBL-coated Kapton in circuit with a LED bulb and power source. (**b**) Spraying of NWs onto 12" x 12" Kapton by manually mobilizing the nozzle. (**c**) Large-area film coated Kapton substrates. (**d**) Photograph of a drone wing, part of which is masked with Kapton substrates, inset: the cross-sectional photograph of the drone wing. (**e**) SEM image of the drone wing. (**f**) Measurement of conductance in the uncoated area of the drone wing. (**g**) Photograph of the curvilinear drone wing after the deposition of (PEI/NWs). (**h**) SEM image of the drone wing after the deposition of (PEI/NWs). (**i**) Measurement of conductance of the drone wing in the area coated with (PEI/NWs).

## 3. CONCLUSIONS

GT can be successfully applied to materials made by LBL as an inherently additive, substrate-agnostic, and component-tolerant materials engineering tool. Integration of GT and LBL uncovers a massive materials design space due to its suitability for a multitude of nanoscale components with realistic size and shape distributions. The predictive capabilities offered by GT will allow navigation of this vast parameter space for efficient design of next-generation high-performance composites. We have shown how LBL experimental toolbox lends itself to GT-driven property predictions for samples of realistic sizes opening the possibility for the investigation of arbitrarily large systems. In future, we expect that increasingly data-driven GT approaches may additionally leverage AI, for an even more efficient composite engineering.

## 4. Experimental Section

*Chemicals:* Poly(ethyleneimine) ($M_n \approx 60{,}000$) and GO were purchased from Sigma-Aldrich (St. Louis, MO). Silver NWs ($d$ = 30 nm, $L$ = 20 – 30 nm) were purchased from Novarials (Woburn, MA). Microscope glass slides were purchased from Thermo Fisher Scientific (Waltham, MA). Other substrates including Kapton sheets, cerium-doped glass slides and fused silica were provided by Boeing (Huntington Beach and El Segundo, CA). All chemicals were used as received without further purification.

*Fabrication of NW multilayer films:* A substrate was washed in ethanol and then water before being cleaned for 10 min in a plasma cleaner (Evactron 25; XEI Scientific, Redwood City, CA). LBL deposition was realized using a spray system as described elsewhere[23,24]. A PEI solution ($C$ = 2.5 mg mL$^{-1}$ in water) was sprayed to the cleaned substrate for 30 s by pumping it to the spray nozzle with the liquid pump at a given speed ($F_L$ = 20 mL min$^{-1}$). Air was fed to the spraying nozzle simultaneously at a given flow rate ($F_A$ = 10 L min$^{-1}$). The sample was then rinsed by water spray for 10s and dried with compressed air. A suspension of NWs ($C$ = 0.2 mg mL$^{-1}$ in water) was then sprayed onto the PEI-coated substrate using the same spraying system, with a liquid flow at $F_L$ = 2 mL min$^{-1}$ and an air flow rate at $F_A$ = 2.5 L min$^{-1}$. An isotropic NW network is formed by spraying the suspension perpendicularly (incidence angle $\theta$ = 90°) to the PEI-coated substrate at a distance ($d$) of 5 cm for a time ($t$) of 100 s. The coated substrate was water rinsed for 10 s to remove weakly attached NWs and dried with compressed air. As a result, a conductive film with a structure written as PEI/NWs is fabricated. The same process can be repeated to build-up (PEI/NWs)$_N$ multilayer structures, with up to 10 bilayers in this work. Anisotropic films with aligned NWs were obtained by changing the incidence angle $\theta$ to 10°, with parameters of $C$ = 0.05 mg mL$^{-1}$; $F_L$ = 2 mL min$^{-1}$; $F_A$ = 20 L min$^{-1}$; $d$ = 1 cm; $t$ = 200 s.

*Electrical Characterization:* The sheet resistance of the conductive films deposited on hard substrates (glass slides) was measured by a four-point probe (SP4; Signatone, Gilroy, CA), powered by a source meter (Model 2470; Keithley, Cleveland, OH). A current of 1 mA was applied and the voltage was measured and recorded. The sheet resistance was calculated using the following equation

$$R_s = (\pi/\ln 2) \times \Delta V/I \qquad \text{Equation 3}$$

where $R_s$ is the sheet resistance, $I = 1$ mA is the applied current, $\Delta V$ is the measured voltage. The result for each is an average of 10 measurements over the entire surface. The sheet resistance variation with respect to temperature (25 °C - 90 °C) was assessed by positioning the samples atop a flexible strip heater (Tempco, Farmington Hills, MI), which had its temperature controlled by a controller (SDX; Briskheat, Columbus, OH). The electrical resistance of the conductive films, deposited on soft substrates (Kapton sheets), was measured by applying a DC voltage across both ends of the 2 cm × 2 cm membrane. An averaged *I-U* curve was generated by sweeping the voltage back and forth from -1 V to 1 V for a total of 10 cycles. The current-carrying capacity was determined by gradually increasing the voltage in increments of 1 V until failure occurred within 1 minute.

*Optical Characterization:* The transmittance/absorbance was measured with an 8453 UV-Vis spectrometer (Agilent, Santa Clara, CA).

*Electron Microscopy:* Scanning electron microscopic (SEM) images of the samples were collected with Nova 200 NanoLab microscope (Thermo Fisher Scientific, Waltham, MA).

*Atomic Force Microscopy:* A Bruker Dimension Icon (Billerica, MA) was used. Silicon nitride probes (ScanAsyst Tapping mode AFM) were purchased from Bruker, and imaging was performed in ambient conditions. Images were obtained at a scan rate of 1 Hz, and 512 points per line. Bruker NanoScope Analysis 1.5 software was used to process the images to flatten each line individually, and to remove tilt and bow. After plane fitting, the images were analyzed for $R_q$, to obtain the surface roughness.

*Mechanical characterization*: (PEI/NWs)$_N$ films were deposited on a PDMS strip presteched with 10% strain. Upon release, buckling patterns were formed owing to the differential stiffness of the NW films and the substrate underneath. The elastic modulus of the film was calculated using equation 4.

$$\frac{E_f}{(1-v_f^2)} = \frac{3E_s}{(1-v_s^2)}(\frac{d}{2\pi h})^3 \qquad \text{Equation 4}$$

where $E_s$ is the elastic modulus of the PDMS substrate (measured to be 2.97 MPa using nanoindentation), $v_s$ (0.5) is the Poisson's ratio of PDMS, $d$ is the wavelength of the buckling measured by AFM, $h$ is the thickness of the NW film (92 nm for the 10-bilayer structure), $v_f$ is the Poission's ratio of the NW film (0.37, Poisson's ratio of bulk silver was used), and $E_f$ is the calculated elastic modulus of the NW film.

Buckling experiments were performed on (PEI/NWs)$_N$ films with $N$ = 1, 2, 3, 5 and 10. The AFM images and the buckling wavelength are given in Figure S7. Elastic modulus was calculated for $N$ = 10 only but not the thinner films as their thickness was too small to be accurately measured from the cross-sectional SEM images.

Elastic modulus of PDMS was measured using nanoindentation on Hysitron TI-950 Nanoindenter (Bruker, USA). A 50-um spherical diamond probe was used under displacement control mode with a displacement of 1 μm.

*Polarized terahertz time-domain spectroscopy (THz-TDS):* To measure the effect of NW on THz pulse, THz-TDS with 3 polarizers was used. THz pulses were generated and detected using photoconductive antennas (PCA, Tera15-FC, Menlo Systems) pumped and probed by 1560 nm pulsed laser in a standard THz-TDS system (TeraSmart, Menlo Systems). Four convex lenses (TPX50, Menlo Systems) were used to generate collimated and focused THz beam, and the sample was positioned at the focus point on a motorized XY stage that enables imaging of the entire sample. Unlike conventional THz-TDS system where the polarization direction of the PCAs align together, PCAs in our polarized THz-TDS system align perpendicularly, so that the emitted THz beam is linearly polarized horizontally to the optical table (defined as the *x* axis) while the detector is measuring linearly polarized THz beam vertical to the optical table (defined as the *y* axis). To ensure the polarization state of these PCAs and neglect all artifacts, fixed wire grid polarizers (P1 and P3) were positioned in front of each PCAs with the same polarization direction regarding to each PCAs. Another polarizer (P2) was placed in between the sample and P3 and was rotated throughout the measurement to acquire the complete polarization state of the transmitted THz beam. Three different measurements were conducted for each sample regarding with the angle of polarization axis of P2: 0°, +45°, -45° with the *y*-axis. The *y*-component of the transmitted THz beam $E_y(t)$ was retrieved by measuring with 0° of P2 and the x-component $E_x(t)$ was retrieved by the subtraction of measured E-field with +45° and -45° of P2 ($E_x(t) = E_{+45°}(t) - E_{-45°}(t)$). Complex frequency-domain signals were then extracted with fast Fourier transform (FFT) of these time-domain signals ($\tilde{E}_{x,y} = \tilde{E}_{x,y}(\omega) = FFT\{E_{x,y}(t)\}$).

*THz Calculation:* THz absorption (TA) was obtained by first calculating the transmittance spectra of the sample and reference from the x- and y-component electric fields.

$$T_s = \sqrt{|\tilde{E}_x|^2 + |\tilde{E}_y|^2} \qquad \text{Equation 5}$$

$$T_{ref} = \sqrt{|\tilde{E}_{x,ref}|^2 + |\tilde{E}_{y,ref}|^2} \qquad \text{Equation 6}$$

$T_s$ is the transmittance spectra of the NW on the PVC substrate, and $T_{ref}$ is the transmittance spectra of the sample holding paper as a reference. Then, TA can be calculated by

$$TA \propto -\ln(T_s/T_{ref}) \qquad \text{Equation 7}$$

To obtain THz optical rotation dispersion (TORD), four Stokes parameters should be calculated with the measured transmitted THz signals with

$$S_0 = \tilde{E}_x \tilde{E}_x^* + \tilde{E}_y \tilde{E}_y^* \qquad \text{Equation 8}$$
$$S_1 = \tilde{E}_x \tilde{E}_x^* - \tilde{E}_y \tilde{E}_y^* \qquad \text{Equation 9}$$
$$S_2 = \tilde{E}_x \tilde{E}_y^* + \tilde{E}_y \tilde{E}_x^* \qquad \text{Equation 10}$$
$$S_3 = i(\tilde{E}_x \tilde{E}_y^* - \tilde{E}_y \tilde{E}_x^*) \qquad \text{Equation 11}$$

With the Stokes parameters, TORD can be calculated by

$$TORD = \frac{1}{2}\tan^{-1}\left(\frac{S_2}{S_1}\right), -\frac{\pi}{2} \leq TORD \leq \frac{\pi}{2} \qquad \text{Equation 12}$$

*Graph theoretic characterization:* The mathematical representation of a graph is given by the *adjacency matrix*, **A**, whose elements are

$$A_{ij} = \begin{cases} R_{ij}^{-1}, \text{if edge between nodes } i \text{ and } j \\ 0, otherwise \end{cases} \qquad \text{Equation 13}$$

Where $R_{ij}$ is the electrical resistance of the wire segment joining nodes $i$ and $j$, given by Pouillet's Law:

$$R_{ij} = 2R_J + \rho_{Ag} \frac{l_{ij}}{a_{ij}} \qquad \text{Equation 14}$$

where $R_J$ is a constant resistance associated with the intersection of NWs (the factor of 2 resulting from 2 wire intersections per edge), $\rho_{Ag}$ is the resistivity of Ag NWs)[46], $l_{ij}$ is the length of the edge connecting nodes $i$ and $j$, and $a_{ij}$ is its cross-sectional area. To extract **A** from an SEM, our open-source package StructuralGT (SGT)[32] is used. SGT extracts graphs from skeletons which were estimated from a binarized image of the original SEM. The binarized image is obtained from the raw image using a user-specified set of image processing options. $R_{ij}$ terms for each NW are estimated from material property data and their geometric dimensions measured from the image. With the extracted **A**, we may write electrical circuitry laws in graph theoretic terms. Namely, Kirchhoff's law and Ohm's law may be written as

$$\sum_j [A_{ij}(V_i - V_j)] - I_i = 0 \qquad \text{Equation 15}$$

where $V_i$ is the voltage at node $i$, $I_i$ is any net current which flows through node $i$. In the context of NW films, $I_i$ is $(-1) / N_{source}$ for all nodes that intersect with the left electrode, $(-1) / N_{sink}$ for all nodes that intersect with the right electrode, and 0 for all other nodes. $N_{source}$ and $N_{sink}$ are the number of left and right electrode connections, respectively, thus ensuring charge is conserved. Equation 15 may also be written as

$$\sum_j [(\delta_{ij} k_i - A_{ij}) V_j] - I_i = 0 \qquad \text{Equation 16}$$

Where $\delta_{ij}$ is the Kroenecker delta and $k_i$ is the *degree of the node*: the number of edges connected to it, equal to $\sum_j A_{ij}$. This can be written in matrix form as[27]

$$\mathbf{LV} = \mathbf{I} \qquad \text{Equation 17}$$

Where we have used the definition of the *graph Laplacian* $\mathbf{L} = (\delta_{ij} k_i - A_{ij})$. By noting that every row of **L** sums to 0, it is always singular, and hence there is no inverse (and hence no single solution to Equation 17). Physically, this singularity is a result of the arbitrariness of choice of reference potential. By constraining our reference potential such that the average voltage is 0, the solution is given by[47]

$$\mathbf{V} = \mathbf{L}^\dagger \mathbf{I} \qquad \text{Equation 18}$$

where $\mathbf{L}^\dagger$ is the *pseudoinverse* of $\mathbf{L}$. (The pseudoinverse provides the best fit to singular linear systems *i.e.,* in this case, there are infinitely many solutions to the voltage distribution, $\mathbf{V}$, all within an additive constant). Furthermore, the network between any pair of non-adjacent nodes may be coarse grained to a single edge, for which the effective resistance is given by

$$\Omega_{kl} = L_{kk}^\dagger + L_{ll}^\dagger - 2L_{kl}^\dagger \qquad \text{Equation 19}$$

where $L_{kl}^\dagger$ is the $kl$th element of $\mathbf{L}^\dagger$. $\Omega_{kl}$ is the effective resistance between nodes $k$ and $l$, defined as

$$\Omega_{kl} \equiv (V_k - V_l)/I_{kl} \qquad \text{Equation 20}$$

When calculating the resistance of an NW film, a current source and sink node are added at opposite ends of the network. Nodes along the length segment located on the opposite ends of the sample are attached to either the source or sink nodes. When $k$ and $l$ correspond to the source and sink, $\Omega_{kl}$ is the film's resistance, $\Omega_{eff}$. In our work, we fit values of $\Omega_{eff}$ to experimental data *via* the parameter $R_J$, the resistance associated with each NW junction. We show the model's predictive capability by confirming that the same $R_J$ value reproduces experimental $\Omega_{eff}$ values at different experimental conditions. Although the constant $R_J$ assumption is not strictly true,[48] in the SI we show that it doesn't introduce significant inaccuracies for sheet resistance predictions. A final quantity of interest is the conductive selectivity, $Y_{pred}$, which is used to assess how selective a film may be to charge transport, depending on the axis along which a potential difference is applied. It is defined as

$$Y_{pred} \equiv \Omega_{eff}/\Omega_{eff}^\perp \qquad \text{Equation 21}$$

Where $\Omega_{eff}^\perp$ is the film's effective resistance when a potential difference is applied along a direction perpendicular to the direction applied in obtaining $\Omega_{eff}$. A point of interest in obtaining $\Omega_{eff}$ from $A$ is what value(s) should be assigned to $R_J$, as this is the only parameter which can be neither easily obtained from SEM nor material property data. In this work, we fit a constant value of $R_J$ to experimental data and confirm the model's ability to extrapolate to new samples.

# Acknowledgments

W.W. and A.K. contributed equally to this work. The authors are grateful for the National Science Foundation and the Air Force Office of Scientific Research for support of this work. The authors acknowledge the Michigan Center for Materials Characterization (MC$^2$) for use of the instruments and staff assistance.

s

# References

[1]     J. D. Swalen, *Science* **1990**, *249*, 305.

[2]     M. Mrksich, *Chem. Soc. Rev.* **2000**, *29*, 267.

[3]     P. M. Martin, *Handbook of Deposition Technologies for Films and Coatings: Science, Applications and Technology*, Elsevier, **2010**.

[4]     D. M. Mattox, *Handbook of Physical Vapor Deposition (PVD) Processing*, William Andrew, **2010**.

[5]     Z. Cai, B. Liu, X. Zou, H.-M. Cheng, *Chem. Rev.* **2018**, *118*, 6091.

[6]     G. Decher, *Science* **1997**, *277*, 1232.

[7]     J. J. Richardson, J. Cui, M. Björnmalm, J. A. Braunger, H. Ejima, F. Caruso, *Chem. Rev.* **2016**, *116*, 14828.

[8]     S. Zhao, F. Caruso, L. Dähne, G. Decher, B. G. De Geest, J. Fan, N. Feliu, Y. Gogotsi, P. T. Hammond, M. C. Hersam, A. Khademhosseini, N. Kotov, S. Leporatti, Y. Li, F. Lisdat, L. M. Liz-Marzán, S. Moya, P. Mulvaney, A. L. Rogach, S. Roy, D. G. Shchukin, A. G. Skirtach, M. M. Stevens, G. B. Sukhorukov, P. S. Weiss, Z. Yue, D. Zhu, W. J. Parak, *ACS Nano* **2019**, *13*, 6151.

[9]     S. Srivastava, N. A. Kotov, *Acc. Chem. Res.* **2008**, *41*, 1831.

[10]    J. J. Richardson, M. Björnmalm, F. Caruso, *Science* **2015**, *348*, aaa2491.

[11]    P. Podsiadlo, A. K. Kaushik, E. M. Arruda, A. M. Waas, B. S. Shim, J. Xu, H. Nandivada, B. G. Pumplin, J. Lahann, A. Ramamoorthy, N. A. Kotov, *Science* **2007**, *318*, 80.

[12]    Z. Tang, N. A. Kotov, S. Magonov, B. Ozturk, *Nature Mater* **2003**, *2*, 413.

[13]    R. Merindol, S. Diabang, O. Felix, T. Roland, C. Gauthier, G. Decher, *ACS Nano* **2015**, *9*, 1127.

[14]    B. S. Shim, Z. Tang, M. P. Morabito, A. Agarwal, H. Hong, N. A. Kotov, *Chem. Mater.* **2007**, *19*, 5467.

[15]    K. J. Loh, J. Kim, J. P. Lynch, N. W. S. Kam, N. A. Kotov, *Smart Mater. Struct.* **2007**, *16*, 429.

[16]    H. Ledford, *Nature* **2015**, *525*, 308.

[17]    O. du Roure, A. Lindner, E. N. Nazockdast, M. J. Shelley, *Annual Review of Fluid Mechanics* **2019**, *51*, 539.

[18]    W. Jiang, Z. Qu, P. Kumar, D. Vecchio, Y. Wang, Y. Ma, J. H. Bahng, K. Bernardino, W. R. Gomes, F. M. Colombari, A. Lozada-Blanco, M. Veksler, E. Marino, A. Simon, C. Murray, S. R. Muniz, A. F. de Moura, N. A. Kotov, *Science* **2020**, *368*, 642.

[19]    J. B. Schlenoff, S. T. Dubas, T. Farhat, *Langmuir* **2000**, *16*, 9968.

[20]    H. Hu, S. Sekar, W. Wu, Y. Battie, V. Lemaire, O. Arteaga, L. V. Poulikakos, D. J. Norris, H. Giessen, G. Decher, M. Pauly, *ACS Nano* **2021**, *15*, 13653.

[21]    S. Sekar, V. Lemaire, H. Hu, G. Decher, M. Pauly, *Faraday Discuss.* **2016**, *191*, 373.

[22]    R. Blell, X. Lin, T. Lindström, M. Ankerfors, M. Pauly, O. Felix, G. Decher, *ACS Nano* **2017**, *11*, 84.

[23]    W. Wu, Y. Battie, V. Lemaire, G. Decher, M. Pauly, *Nano Lett.* **2021**, *21*, 8298.

[24]    W. Wu, Y. Battie, C. Genet, T. W. Ebbesen, G. Decher, M. Pauly, *Advanced Optical Materials* **2023**, *11*, 2202831.

[25]    D. J. Watts, S. H. Strogatz, *Nature* **1998**, *393*, 440.

[26]    Network Science, National Academies Press, Washington, D.C., **2005**.

[27]    M. Newman, *Networks*, Second Edition, New to this Edition:, Second Edition, New to this Edition:, Oxford University Press, Oxford, New York, **2018**.

[28]    M. Wang, D. Vecchio, C. Wang, A. Emre, X. Xiao, Z. Jiang, P. Bogdan, Y. Huang, N. A. Kotov, *Science Robotics* **2020**, *5*, eaba1912.

[29]    D. A. Vecchio, M. D. Hammig, X. Xiao, A. Saha, P. Bogdan, N. A. Kotov, *Advanced Materials* **2022**, *34*, 2201313.

[30]	H. Zhang, D. Vecchio, A. Emre, S. Rahmani, C. Cheng, J. Zhu, A. C. Misra, J. Lahann, N. A. Kotov, *MRS Bulletin* **2021**, *46*, 576.

[31]	N. Kotov, M. Wang, K. Whishant, V. Cecen, L. Zhao, Z. Zhong, L. Liu, Y. Huang, *Topometric Design of Reticulated Nanofiber Composites for Lithium-Sulfur Batteries*, In Review, **2023**.

[32]	D. A. Vecchio, S. H. Mahler, M. D. Hammig, N. A. Kotov, *ACS Nano* **2021**, *15*, 12847.

[33]	M. Jagota, N. Tansu, *Sci Rep* **2015**, *5*, 10219.

[34]	R. Benda, E. Cancès, B. Lebental, *Journal of Applied Physics* **2019**, *126*, 044306.

[35]	M. Jagota, I. Scheinfeld, *Phys. Rev. E* **2020**, *101*, 012304.

[36]	A. Mohammadpour-Haratbar, Y. Zare, K. Y. Rhee, *Sci Rep* **2023**, *13*, 5.

[37]	P. Ercius, O. Alaidi, M. J. Rames, G. Ren, *Advanced Materials* **2015**, *27*, 5638.

[38]	Z. Cheng, L. Liu, S. Xu, M. Lu, X. Wang, *Sci Rep* **2015**, *5*, 10718.

[39]	C. M. Stafford, C. Harrison, K. L. Beers, A. Karim, E. J. Amis, M. R. VanLandingham, H.-C. Kim, W. Volksen, R. D. Miller, E. E. Simonyi, *Nature Mater* **2004**, *3*, 545.

[40]	S. Wang, X. Zhang, X. Wu, C. Lu, *Soft Matter* **2016**, *12*, 845.

[41]	M. Pournajar, M. Zaiser, P. Moretti, *Sci Rep* **2022**, *12*, 11814.

[42]	J. Hwang, H. Sohn, S. H. Lee, *Sci Rep* **2018**, *8*, 16617.

[43]	D. Chen, Y. Zhang, T. Bessho, J. Sang, H. Hirahara, K. Mori, Z. Kang, *Chemical Engineering Journal* **2016**, *303*, 100.

[44]	K. C. Krogman, J. L. Lowery, N. S. Zacharia, G. C. Rutledge, P. T. Hammond, *Nature Mater* **2009**, *8*, 512.

[45]	A. P. R. Johnston, C. Cortez, A. S. Angelatos, F. Caruso, *Current Opinion in Colloid & Interface Science* **2006**, *11*, 203.

[46]	A. Bid, A. Bora, A. K. Raychaudhuri, *Phys. Rev. B* **2006**, *74*, 035426.

[47]	P. Van Mieghem, K. Devriendt, H. Cetinay, *Phys. Rev. E* **2017**, *96*, 032311.

[48]	A. T. Bellew, H. G. Manning, C. Gomes da Rocha, M. S. Ferreira, J. J. Boland, *ACS Nano* **2015**, *9*, 11422.

Spray-assisted layer-by-layer assembly is an excellent strategy for the fabrication of multilayered thin films made from silver NWs and other nanomaterials on flexible and curvilinear substrates. The films exhibit a set of excellent properties, which are analyzed by graph theory. Findings presented in this work demonstrate a possibility of rational design and fast fabrication of multifunctional nanocomposite coatings.

Wenbing Wu, Alain Kadar, Sang Hyun Lee, Bum Chul Park, Jeffery E. Raymond, Thomas K. Tsotsis, Carlos E. S. Cesnik, Sharon C. Glotzer,* Valerie Goss,* Nicholas A. Kotov*

**Layer-by-layer Assembled NW Networks Enable Graph Theoretical Design of Multifunctional Coatings**

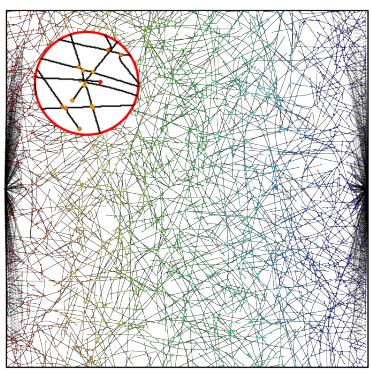